# Structure and Dynamics of a Lattice of Tetragonal Germanates $R_2Ge_2O_7$ (R = Tb–Lu, Y): Ab Initio Calculation

V. S. Ryumshin[a,] * and V. A. Chernyshev[a]

[a] *Ural Federal University, Yekaterinburg, 620002 Russia*

*e-mail: krios_two@mail.ru*

**Abstract**—The crystal structure, phonon spectrum, and elastic constants of a series of rare-earth germanates (including yttrium germanate $R_2Ge_2O_7$ (R = Tb–Lu, Y)) with a tetragonal structure have been ab initio calculated within the density functional theory. The frequencies and types of fundamental vibrations and the intensities of IR and Raman modes are determined. The degrees of participation of ions in each mode are determined by analyzing the displacement vectors obtained as a result of the ab initio calculations. The calculations have been performed for the first time; there are no corresponding experimental data for the entire series of compounds (except for the IR and Raman spectra of yttrium germanate). The performed calculations made it possible to interpret and supplement the known data in the literature on IR and Raman spectra of yttrium germanate $Y_2Ge_2O_7$.

Much attention is paid to $R_2Ge_2O_7$ crystals (R is a rare earth ion) because of the variety of their properties [1–5]. They are crystallized into different structure types [1, 6]. The crystal structure of these compounds has been thoroughly studied [2, 7]; however, there are no experimental data on the phonon spectra and elastic constants for series rare-earth germanates. Rare-earth germanates are promising as scintillators [8]; yttrium germanate $Y_2Ge_2O_7$ is used as a matrix for doping by rare-earth ions [9, 10]. IR and Raman spectra of $Y_2Ge_2O_7$ were measured in [11]; however, the low-frequency spectral region (up to 200 cm$^{-1}$), which contains modes with the maximum degree of participation of yttrium ions, was not investigated. Thus, experimental data on the germanate modes with the maximum degree of participation of rare-earth ions is absent. Only 17 out of 48 IR-active modes and only 20 out of 81 Raman modes were determined in [11]. The measurements in [11] were performed using a polycrystalline sample, and the mode types were not experimentally determined. Thus, the known experimental data on IR and Raman spectra of yttrium germanate should be complemented and interpreted. Experimental data for other representatives of the series (R = Tb–Lu) are also absent. Specific features of the $Y_2Ge_2O_7$ structure were previously discussed in [12], where it was noted that yttrium germanate is crystallized into the sp. gr. $P4_32_12$ (it makes an enantiomorphic pair with the group $P4_12_12$, into which the entire series R = Tb–Lu is crystallized). Thus, the crystal structure of yttrium germanate is similar to the specular reflection of the structure of rare-earth germanates $R_2Ge_2O_7$ (R = Tb–Lu).

It seems actual to ab initio investigate the phonon spectrum of rare-earth germanates $R_2Ge_2O_7$ (R = Tb–Lu, Y) with a tetragonal structure. The purpose of this study was to analyze within the unified ab initio approach the crystal structure, phonon spectrum, and elastic properties of a series of germanates $R_2Ge_2O_7$ (R = Tb–Lu, Y).

## 1. CALCULATION TECHNIQUES

Calculations were performed within the density functional theory and the approach of linear combination of atomic orbitals (MO LCAO). The hybrid functional PBE0 [13], which takes into account the contribution from nonlocal exchange (in the Hartree–Fock formalism), and non-dynamic correlations were used [14]. The calculations were carried out using the CRYSTAL17 program [15], which is designed for simulating periodic structures. Note that the phonon spectrum of germanates with a pyrochlore structure was calculated previously within the same approach [16].

Ge and O were described using the full-electron basis sets [17, 18]. The inner electron shells of rare-earth Tb–Lu and Y ions were described using the pseudopotentials from [19, 20]. Test calculations were performed by an example of $Gd_2Ge_2O_7$ with a pyrochlore structure using two pseudopotentials: ECP53MWB-1 [19], which replaces inner shells up to the 4$f$ shell inclusively, and ECP28MWB_SEG [21], which replaces inner shells only up to the 3$d$ shell and leaves the 4$f$ shell in the valence balance (i.e., it enters the valence basis set in this case and is described directly). The results are compared in Table 1, which also contains the experimental data from [22]. Parameter ρ = (Gd–O2/Gd–O1) characterizes the degree of distortion of the octahedron with a rare-earth ion at the center and oxygen ions at the vertices. Reproduction of this parameter may serve as an estimate of adequacy of the used basis sets.

**Table 1.** Gd$_2$Ge$_2$O$_7$ lattice constant

| Parameters | Calculation (ECP53) | Calculation (ECP28) | Experiment [22] |
|---|---|---|---|
| $a$, Å | 10.053 | 9.991 | 9.999 |
| Gd–O1, Å | 2.505 | 2.482 | 2.535 |
| Gd–O2, Å | 2.177 | 2.163 | 2.165 |
| ρ | 0.869 | 0.871 | 0.854 |

The use of the "short" pseudopotential (i.e., explicit description of the 4$f$ shell using a basis set) somewhat improves reproduction of the lattice constant, but increases significantly the computing time, which is a critical factor for simulation of the low-symmetry structures under consideration.

In this study, inner shells of rare-earth ions were described using quasi-relativistic ECPnMWB pseudopotentials, where ECP is for "effective core potential", WB is for "quasi-relativistic", and n is the number of inner electrons replaced by the pseudopotential ($n$ = 54 for Tb, 55 for Dy, etc.) [19, 23]. Thus, inner shells of a rare-earth ion (including 4$f$) were replaced by the pseudopotential. The outer shells ($5s^25p^6$), involved in chemical bonds, were described using the ECPnMWB-I valence basis sets [23–25], which contain radial functions of the $s$, $p$, and $d$ type. Pseudopotential ECP28MWB with a corresponding basis set was used for yttrium. The pseudopotentials and valence basis sets are available at the Stuttgart Group website [26]. Gaussian primitives with the exponential factors below 0.1 were removed from the valence basis sets, which is characteristic of periodic calculations.

The calculation algorithm was as follows. First, the crystal structure was optimized: the lattice constants and ionic coordinates in a cell were determined. The phonon spectrum (at the Γ point) or elastic constants were calculated for the obtained crystal structure corresponding to the energy minimum.

Integration over the Brillouin zone was performed using the Monkhorst–Pack scheme with a $k$-point grid 4 × 4 × 4 in size. This grid was chosen based on the test calculations carried out for erbium germanate with a tetragonal structure (Er$_2$Ge$_2$O$_7$) (Table 2).

**Table 2.** Calculation of the crystal structure of tetragonal Er$_2$Ge$_2$O$_7$ using different grids

| Grid | | Lattice constant, Å | SCF energy, Hartree |
|---|---|---|---|
| 2 × 2 × 2 | $a$ | 6.84380243 | −19027.960230 |
| | $c$ | 12.43678912 | |
| 4 × 4 × 4 | $a$ | 6.84377960 | −19027.960928 |
| | $c$ | 12.43704974 | |
| 6 × 6 × 6 | $a$ | 6.84378081 | −19027.960928 |
| | $c$ | 12.43705748 | |
| 8 × 8 × 8 | $a$ | 6.84378073 | −19027.960928 |
| | $c$ | 12.43707140 | |

It should be noted that the energy of self-consistent field (SCF) was calculated with an accuracy of $10^{-6}$ Hartree. (Accuracy of calculating the two-electron integrals was no less than $10^{-7}$ Hartree.) As follows from the calculations, the 4 × 4 × 4 grid is quite sufficient. At a finer grid, the SCF energies are retained accurate to the sixth decimal place. The lattice constants at a finer grid change at the fifth decimal place. This calculation accuracy can be assumed quite acceptable, because the difference in the calculated and experimental lattice constants is within 0.1 Å (a typical value for ab initio calculations).

## 2. CRYSTAL STRUCTURE

Rare-earth germanates R$_2$Ge$_2$O$_7$ (R = Tb–Lu) have a tetragonal crystal structure and belong to the sp. gr. $P4_12_12$ (92). A unit cell (Fig. 1) contains $Z$ = 4 formula units and 44 atoms. R and Ge ions are surrounded by seven and four O atoms, respectively (Figs. 2, 3). The results of calculating the crystal structure are reported in Tables 3–6.

The ionic coordinates in a cell of Er$_2$Ge$_2$O$_7$ were determined by an X-ray diffraction (XRD) analysis. The calculation results and experimental data are compared in Table 4.

The results of calculating the crystal structure are in good agreement with the known experimental data. The lattice constants and bond lengths decrease along the Tb–Lu series, which corresponds to lanthanide compression. The R–O bond length shortens by 0.06–0.09 Å as moving from Tb to Lu. Note that the Ge–O bond length (O = O2, O3, and O4) shortens by 0.008–0.027 Å and the Ge–O1 bond length decreases by 0.08 Å. Thus, replacement of the rare-earth ion barely affects the Ge–O bonds (O = O2, O3, and O4), while the Ge–O1 bond changes similarly to the R–O bond. Ion O1, in turn, bounds the centers of GeO$_4$ tetrahedra and is not the nearest neighbor of the rare earth ion.

## 3. VIBRATIONAL SPECTRUM

Germanates R$_2$Ge$_2$O$_7$ (R = Tb–Lu, Y) with a tetragonal structure are characterized by the following phonon modes at the Γ point:

$$\Gamma = 16A_1(R) + 17A_2(IR) + 17B_1(R) + 16B_2(R) + 33E(R, IR),$$

where R and IR are Raman and infra-red modes, respectively. Two modes are translational: one A$_2$ and one E (a doubly degenerate mode). The results of calculating the phonon modes at the Γ point, their types, and the IR and Raman intensities are listed in Tables 7 and 8. A change in the frequencies in the Tb–Lu series is shown in Fig. 3.

A comparison of the calculation results with the only known experimental data (for Y$_2$Ge$_2$O$_7$) demonstrates that they are in good agreement (Figs. 4, 5). The intensities of the Raman modes are calculated for the excitation wavelength of 488 nm and $T = 300$ K, which corresponds to the experimental conditions. The spectrum was simulated using Lorenzians with a half-width of 7 cm$^{-1}$. In [11], 20 Raman and 17 IR modes were experimentally found (the Raman mode at 790 cm$^{-1}$ was determined by approximation). The mode types were not determined. The calculations supplement the experimental data [11], which do not include the low-frequency spectral range. An yttrium ion (a rare earth ion in the case of R = Tb–Lu) participates most actively in vibrations in specifically this range. The mode types indicated in Figs. 4 and 5 are obtained from the calculations.

The degrees of participation of ions in each mode were determined by analyzing the displacement vectors obtained from the ab initio calculation. The ionic displacements in each vibration for Y$_2$Ge$_2$O$_7$ and Er$_2$Ge$_2$O$_7$ are shown in Figs. 6 and 7, respectively.

According to the calculations, the strongest participations of Y and Ge ions are observed at frequencies 67 cm$^{-1}$ (A$_2$ mode) and 88 cm$^{-1}$ (E mode), respectively. The strongest participation of oxygen ions was predicted in the following modes: E mode with a frequency of 255 cm$^{-1}$ (O1), E mode with a frequency of 216 cm$^{-1}$ (O2 and O3), and A$_2$ mode with a frequency of 134 cm$^{-1}$ (O4). Active participation of Y and Ge ions is observed in the modes with frequencies up to ~300 and ~500 cm$^{-1}$, respectively. An O ion participates actively in all modes.

The strongest IR mode (698 cm$^{-1}$, E mode) mainly involves ions O1 and O2. The strongest mode in the Raman spectrum is A$_1$ with a frequency of 857 cm$^{-1}$,

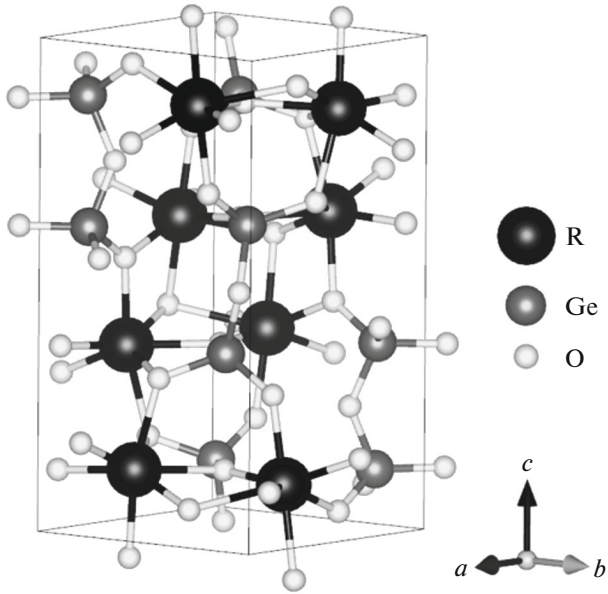

**Fig. 1.** R$_2$Ge$_2$O$_7$ unit cell: tetragonal structure, sp. gr. $P4_12_12$ (92).

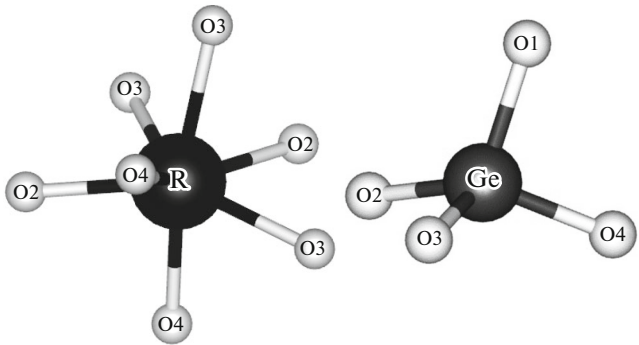

**Fig. 2.** R and Ge ions with their environments.

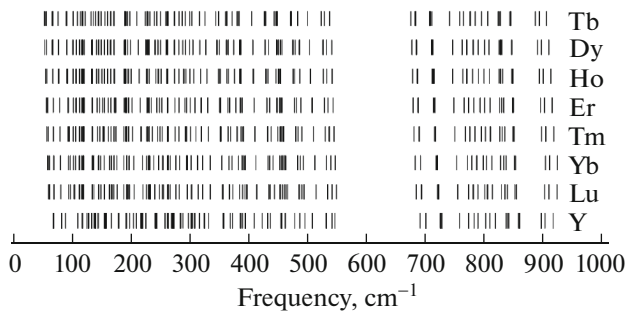

**Fig. 3.** Frequencies of the phonon modes of R$_2$Ge$_2$O$_7$ at the Γ point.

Table 3. R$_2$Ge$_2$O$_7$ lattice constants (R = Tb–Lu, Y), Å

| R | | Calculation | Experiment | | Δ |
|---|---|---|---|---|---|
| Tb | a | 6.9167 | 6.8554 | [7] | 0.0613 |
| | c | 12.5787 | 12.4634 | | 0.1153 |
| Dy | a | 6.8917 | 6.8269 | [27] | 0.0648 |
| | c | 12.5302 | 12.4289 | | 0.1013 |
| Ho | a | 6.8675 | 6.8068 | [27] | 0.0589 |
| | c | 12.4836 | 12.3812 | | 0.1024 |
| Er | a | 6.8438 | 6.7849 | [28] | 0.0589 |
| | c | 12.4371 | 12.3380 | | 0.0991 |
| Tm | a | 6.8232 | 6.7645 | [29] | 0.0587 |
| | c | 12.3953 | 12.2930 | | 0.1023 |
| Yb | a | 6.8011 | 6.7426 | [7] | 0.0585 |
| | c | 12.3496 | 12.2604 | | 0.0892 |
| Lu | a | 6.7845 | 6.7278 | [7] | 0.0567 |
| | c | 12.3140 | 12.2246 | | 0.0894 |
| Y* | a | 6.8196 | 6.8022 | [12] | 0.0174 |
| | c | 12.3955 | 12.3759 | | 0.0196 |

* Calculation for yttrium germanate is performed in the enantiomorphic group $P4_32_12$, which corresponds to the experimental data.

Table 4. Coordinates of Er$_2$Ge$_2$O$_7$ ions, rel. units

| Ion | | Site | Calculation | Experiment [30] |
|---|---|---|---|---|
| Er | x | 8a | 0.8757 | 0.8756 |
| | y | | 0.3521 | 0.3527 |
| | z | | 0.1349 | 0.1355 |
| Ge | x | 8a | 0.9015 | 0.9004 |
| | y | | 0.1526 | 0.1521 |
| | z | | 0.6193 | 0.6196 |
| O1 | x | 4a | 0.8061 | 0.055 |
| | y | | 0.1939 | 0.1945 |
| | z | | 0.7500 | 0.7500 |
| O2 | x | 8a | 0.0786 | 0.0764 |
| | y | | −0.0301 | −0.0331 |
| | z | | 0.6226 | 0.6236 |
| O3 | x | 8a | 0.0647 | 0.0638 |
| | y | | 0.3377 | 0.3355 |
| | z | | 0.5717 | 0.5751 |
| O4 | x | 8a | 0.6850 | 0.6833 |
| | y | | 0.1452 | 0.1449 |
| | z | | 0.5454 | 0.5427 |

in which ions O3 and O4 actively participate. In the second strongest Raman mode with a frequency of 455 cm$^{-1}$ (also A$_1$), ion O4 participates most strongly. The high-frequency range (~700–900 cm$^{-1}$) of both the IR and Raman spectra is due to the dominant participation of an oxygen ion. In the high-frequency E mode (916 cm$^{-1}$), ion O4 participates most actively; however, this mode is weak.

In rare-earth germanates, active participation of rare-earth ions is observed in the low-frequency modes with frequencies up to ~200 cm$^{-1}$ (~400 cm$^{-1}$ for Ge). Oxygen participates in all modes. The frequencies of the strongest modes increase in the series from Tb to Lu within 11 cm$^{-1}$. The highest degree of participation of a rare earth ion in the Tb–Lu series is observed in the low-frequency modes and at a frequency of 103–107 cm$^{-1}$ (E mode). The displacement value decreases with lanthanide compression. A germanium ion is characterized by the largest displacement at a frequency of 138–144 cm$^{-1}$ ($B_1$ mode). Oxygen ions participate in all modes. The strongest participation of O ions in the series was predicted for the following modes: $A_2$ mode at frequencies in the range of 130–136 cm$^{-1}$ (O1), $A_2$ mode at frequencies in the range of 190–195 cm$^{-1}$ (O2), E mode at frequencies in the range of 239–247 cm$^{-1}$ in the Dy–Lu series, $A_2$ mode at a frequency of 344 cm$^{-1}$ for Tb (O3), and E mode at frequencies in the range of 163–167 cm$^{-1}$ (O4). Strong participation of oxygen ions is also observed in the high-frequency spectral range.

Table 5. Ion–ion distances, comparison with the experiment data (in Å)

| Ion | | Er | | Tm | | Y | |
|---|---|---|---|---|---|---|---|
| | | calculation | experiment [30] | calculation | experiment [4] | calculation | experiment [12] |
| R–O | O2 | 2.2310 | 2.195 | 2.2224 | 2.197 | 2.2139 | 2.211 |
| | O4 | 2.2613 | 2.216 | 2.2509 | 2.231 | 2.2499 | 2.245 |
| | O4' | 2.2878 | 2.266 | 2.2756 | 2.252 | 2.2762 | 2.279 |
| | O3 | 2.3002 | 2.278 | 2.2901 | 2.273 | 2.2953 | 2.284 |
| | O2' | 2.3719 | 2.371 | 2.3613 | 2.345 | 2.3800 | 2.373 |
| | O3' | 2.4066 | 2.421 | 2.3924 | 2.362 | 2.3901 | 2.386 |
| | O3" | 2.5511 | 2.516 | 2.5418 | 2.540 | 2.5446 | 2.552 |
| Ge–O | O2 | 1.7420 | 1.733 | 1.7420 | 1.740 | 1.7394 | 1.737 |
| | O4 | 1.7442 | 1.751 | 1.7441 | 1.732 | 1.7407 | 1.733 |
| | O1 | 1.7743 | 1.756 | 1.7729 | 1.751 | 1.7744 | 1.769 |
| | O3 | 1.7894 | 1.754 | 1.7890 | 1.769 | 1.7850 | 1.780 |

**Table 6.** Calculated ion–ion distances in $R_2Ge_2O_7$ (R = Tb–Lu, Y), Å

| Ion | | Tb | Dy | Ho | Er | Tm | Yb | Lu | Y |
|---|---|---|---|---|---|---|---|---|---|
| R–O | O2 | 2.2619 | 2.2513 | 2.2412 | 2.2310 | 2.2224 | 2.2135 | 2.2068 | 2.2139 |
| | O4 | 2.2977 | 2.2853 | 2.2729 | 2.2613 | 2.2509 | 2.2392 | 2.2307 | 2.2499 |
| | O4' | 2.3298 | 2.3151 | 2.3018 | 2.2878 | 2.2756 | 2.2628 | 2.2528 | 2.2762 |
| | O3 | 2.3355 | 2.3231 | 2.3111 | 2.3002 | 2.2901 | 2.2788 | 2.2707 | 2.2953 |
| | O2' | 2.4079 | 2.3957 | 2.3836 | 2.3719 | 2.3613 | 2.3507 | 2.3417 | 2.3800 |
| | O3' | 2.4555 | 2.4385 | 2.4231 | 2.4066 | 2.3924 | 2.3779 | 2.3660 | 2.3901 |
| | O3" | 2.5856 | 2.5739 | 2.5621 | 2.5511 | 2.5418 | 2.5313 | 2.5243 | 2.5446 |
| Ge–O | O2 | 1.7425 | 1.7423 | 1.7419 | 1.7420 | 1.7420 | 1.7417 | 1.7417 | 1.7394 |
| | O4 | 1.7449 | 1.7447 | 1.7445 | 1.7442 | 1.7441 | 1.7437 | 1.7436 | 1.7407 |
| | O1 | 1.7786 | 1.7771 | 1.7756 | 1.7743 | 1.7729 | 1.7716 | 1.7705 | 1.7744 |
| | O3 | 1.7907 | 1.7902 | 1.7900 | 1.7894 | 1.7890 | 1.7884 | 1.7880 | 1.7850 |

## 4. ELASTIC CONSTANTS

The results of calculating elastic constants, bulk modulus, shear modulus (according to Hill), and Vickers hardness are listed in Table 9. Hardness $H_V$ was estimated from formula (1) from [31], where it was successfully used to describe a series of ~40 compounds with ionic and covalent bonds. As was noted in [31], this formula yields the best agreement with the experimental data at hardness more than 5 GPa. In formula (1), $B$ is the bulk modulus and $G$ is the shear modulus.

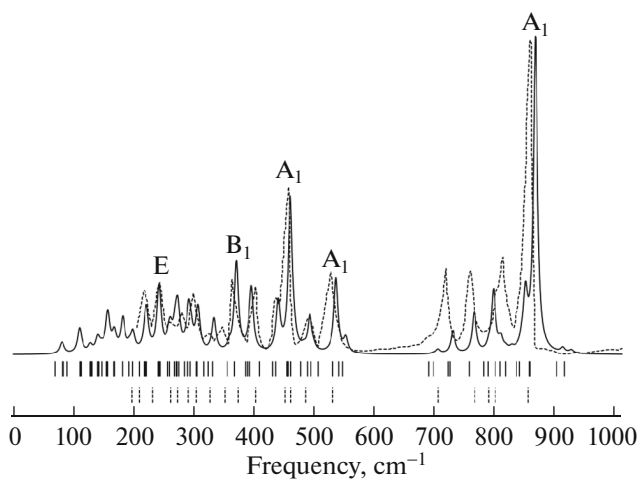

**Fig. 4.** Comparison of the (solid curve) calculated and (dashed curve) experimental [10] Raman spectra of $Y_2Ge_2O_7$.

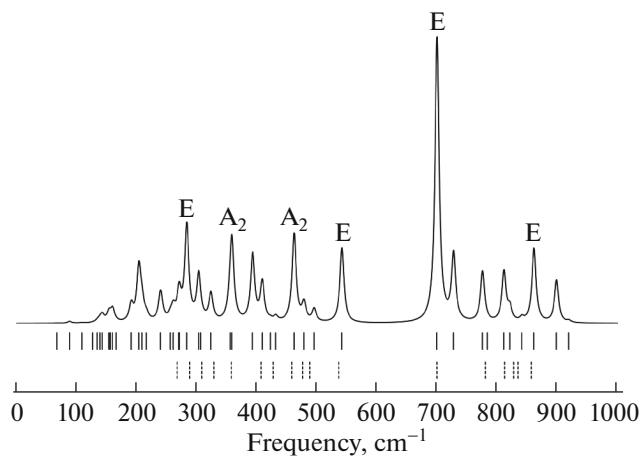

**Fig. 5.** Comparison of the (solid curve) calculated and (dashed curve) experimental [10] IR spectra of $Y_2Ge_2O_7$.

**Table 7.** Phonon modes of $Y_2Ge_2O_7$

| Frequency, cm$^{-1}$ | Type | IR intensity, km/mol | Raman intensity, rel. units | Frequency, cm$^{-1}$ | Type | IR intensity, km/mol | Raman intensity, rel. units |
|---|---|---|---|---|---|---|---|
| 67 | $A_2$ | 5 | — | 314 | $B_2$ | — | 0.02 |
| 68 | $B_1$ | — | 0 | 323 | E | 627 | 0.85 |
| 79 | $A_1$ | — | 22 | 330 | $A_1$ | — | 104 |
| 81 | $B_2$ | — | 17 | 355 | E | 413 | 19 |
| 88 | E | 43 | 0.02 | 357 | $A_2$ | 1697 | — |
| 109 | E | 10 | 73 | 366 | $B_2$ | — | 0.34 |
| 111 | $B_1$ | — | 9 | 367 | $B_1$ | — | 287 |
| 126 | $B_2$ | — | 6 | 384 | $A_1$ | — | 3 |
| 127 | E | 3 | 16 | 388 | $B_2$ | — | 21 |
| 129 | $A_1$ | — | 1.07 | 391 | E | 3 | 192 |
| 134 | $A_2$ | 35 | — | 392 | $A_2$ | 1527 | — |
| 138 | E | 68 | 37 | 408 | E | 891 | 0.92 |
| 140 | $B_1$ | — | 9 | 421 | $A_2$ | 33 | — |
| 142 | $A_2$ | 160 | — | 430 | E | 108 | 0.16 |
| 145 | $B_1$ | — | 14 | 435 | $A_1$ | — | 154 |
| 153 | E | 201 | 14 | 453 | $B_1$ | — | 54 |
| 155.5 | E | 16 | 0.06 | 454 | $B_2$ | — | 13 |
| 155.7 | $A_1$ | — | 113 | 455 | $A_1$ | — | 422 |
| 160 | $A_2$ | 282 | — | 460 | E | 536 | 32 |
| 165.5 | E | 8 | 21 | 461 | $A_2$ | 1496 | — |
| 166.3 | $A_1$ | — | 38 | 477 | E | 407 | 11 |
| 180 | $B_2$ | — | 104 | 487 | $B_1$ | — | 111 |
| 190.2 | $A_2$ | 17 | — | 494 | E | 288 | 11 |
| 190.4 | E | 365 | 0.93 | 506 | $B_2$ | — | 0.46 |
| 190.8 | $B_1$ | — | 15 | 530 | $A_1$ | — | 236 |
| 197 | $B_2$ | — | 58 | 540 | E | 1701 | 0.79 |
| 203 | $A_2$ | 1224 | — | 546 | $B_2$ | — | 44 |
| 208 | E | 339 | 1.28 | 690 | $A_1$ | — | 2 |
| 215.6 | E | 126 | 0.18 | 697.5 | E | 6544 | 11 |
| 216 | $B_1$ | — | 17 | 697.7 | $B_1$ | — | 2 |
| 217.6 | $B_2$ | — | 58 | 722 | $B_2$ | — | 71 |
| 218 | $B_1$ | — | 13 | 724.9 | E | 16 | 0.82 |
| 220 | $A_1$ | — | 67 | 725 | $A_2$ | 1497 | — |
| 239 | E | 682 | 176 | 757 | $B_1$ | — | 130 |
| 240 | $A_1$ | — | 33 | 773 | $A_2$ | 1148 | — |
| 242 | $B_1$ | — | 0.06 | 781 | E | 0.36 | 17 |
| 255 | E | 102 | 29 | 789 | $A_1$ | — | 195 |
| 258 | $B_2$ | — | 58 | 801 | $B_2$ | — | 38 |
| 260 | $A_2$ | 276 | — | 809 | E | 1145 | 9 |
| 266 | $A_1$ | — | 61 | 818 | E | 295 | 10 |
| 269 | E | 349 | 42 | 836 | $B_2$ | — | 19 |
| 270 | E | 318 | 87 | 838 | $A_2$ | 91 | — |
| 273 | $B_1$ | — | 3 | 841 | $B_1$ | — | 167 |
| 282 | E | 2190 | 12.36 | 857 | $A_1$ | — | 1000 |
| 288 | $B_2$ | — | 124 | 858 | E | 1695 | 6 |
| 292 | $A_1$ | — | 46 | 896 | $A_2$ | 966 | — |
| 302 | E | 1048 | 35 | 903 | $B_1$ | — | 15 |
| 304 | $B_1$ | — | 104 | 916 | E | 45 | 10 |
| 306 | $A_2$ | 27 | — | | | | |

**Table 8.** Phonon modes of Er$_2$Ge$_2$O$_7$

| Frequency, cm$^{-1}$ | Type | IR intensity, km/mol | Raman intensity, rel. units | Frequency, cm$^{-1}$ | Type | IR intensity, km/mol | Raman intensity, rel. units |
|---|---|---|---|---|---|---|---|
| 56 | A$_2$ | 5 | – | 308 | B$_2$ | – | 0.18 |
| 58 | B$_1$ | – | 0.78 | 318 | E | 864 | 1 |
| 67.1 | B$_2$ | – | 20 | 330 | A$_1$ | – | 76 |
| 67.2 | A$_1$ | – | 4 | 349 | A$_2$ | 1683 | – |
| 78 | E | 26 | 3 | 352 | E | 359 | 16 |
| 92 | B$_1$ | – | 2 | 363 | B$_2$ | – | 0.03 |
| 93 | E | 0.96 | 71 | 367 | B$_1$ | – | 303 |
| 101 | A$_1$ | – | 8 | 378 | A$_1$ | – | 0.26 |
| 106 | E | 0.26 | 88 | 385 | A$_2$ | 1536 | – |
| 112 | E | 9 | 13 | 387.8 | B$_2$ | – | 31 |
| 116 | B$_2$ | – | 37 | 387.9 | E | 11 | 184 |
| 117 | B$_1$ | – | 40 | 409 | E | 953 | 1 |
| 119 | A$_2$ | 165 | – | 430 | A$_2$ | 131 | – |
| 120 | A$_1$ | – | 72 | 430.4 | E | 53 | 0.03 |
| 133 | A$_2$ | 105 | – | 435 | A$_1$ | – | 198 |
| 134 | E | 5 | 18 | 447 | B$_2$ | – | 12 |
| 141.7 | E | 212 | 2 | 451 | A$_1$ | – | 416 |
| 142.8 | B$_1$ | – | 0.2 | 453 | B$_1$ | – | 80 |
| 146 | E | 92 | 9 | 455 | E | 327 | 37 |
| 151 | A$_2$ | 142 | – | 456 | A$_2$ | 1039 | – |
| 155 | B$_2$ | – | 19 | 477 | E | 760 | 7 |
| 161 | A$_1$ | – | 47 | 481 | B$_1$ | – | 118 |
| 165 | E | 279 | 6 | 488 | E | 496 | 16 |
| 166 | B$_1$ | – | 0.05 | 507 | B$_2$ | – | 3 |
| 171 | A$_2$ | 52 | – | 528 | A$_1$ | – | 236 |
| 173 | B$_2$ | – | 68 | 534 | E | 1509 | 1 |
| 188 | E | 740 | 0 | 543 | B$_2$ | – | 42 |
| 190 | B$_2$ | – | 50 | 679 | A$_1$ | – | 2 |
| 191 | A$_2$ | 1274 | – | 687 | B$_1$ | – | 3 |
| 192 | B$_1$ | – | 31 | 688 | E | 6284 | 11 |
| 195 | A$_1$ | – | 74 | 713 | B$_2$ | – | 77 |
| 201 | E | 28 | 0.12 | 715 | E | 24 | 1 |
| 215 | B$_1$ | – | 8 | 716 | A$_2$ | 1382 | – |
| 226 | B$_1$ | – | 0.12 | 748 | B$_1$ | – | 136 |
| 227 | E | 717 | 58 | 765 | A$_2$ | 1148 | – |
| 229 | B$_2$ | – | 9 | 773 | E | 0.02 | 15 |
| 230 | A$_2$ | 19 | – | 782 | A$_1$ | – | 200 |
| 233 | A$_1$ | – | 42 | 793 | B$_2$ | – | 37 |
| 243 | E | 190 | 67 | 800 | E | 1161 | 8 |
| 247 | E | 233 | 148 | 809 | E | 379 | 8 |
| 252 | A$_1$ | – | 59 | 826 | B$_2$ | – | 20 |
| 261.37 | E | 71 | 43 | 829 | A$_2$ | 98 | – |
| 261.38 | B$_1$ | – | 21 | 833 | B$_1$ | – | 165 |
| 262 | B$_2$ | – | 249 | 848 | A$_1$ | – | 1000 |
| 274 | E | 28 | 42 | 849 | E | 1551 | 6 |
| 281 | E | 1905 | 15 | 895 | A$_2$ | 1087 | – |
| 290 | A$_2$ | 37 | – | 902 | B$_1$ | – | 16 |
| 293 | A$_1$ | – | 71 | 915 | E | 50 | 12 |
| 300 | B$_1$ | – | 69 | | | | |

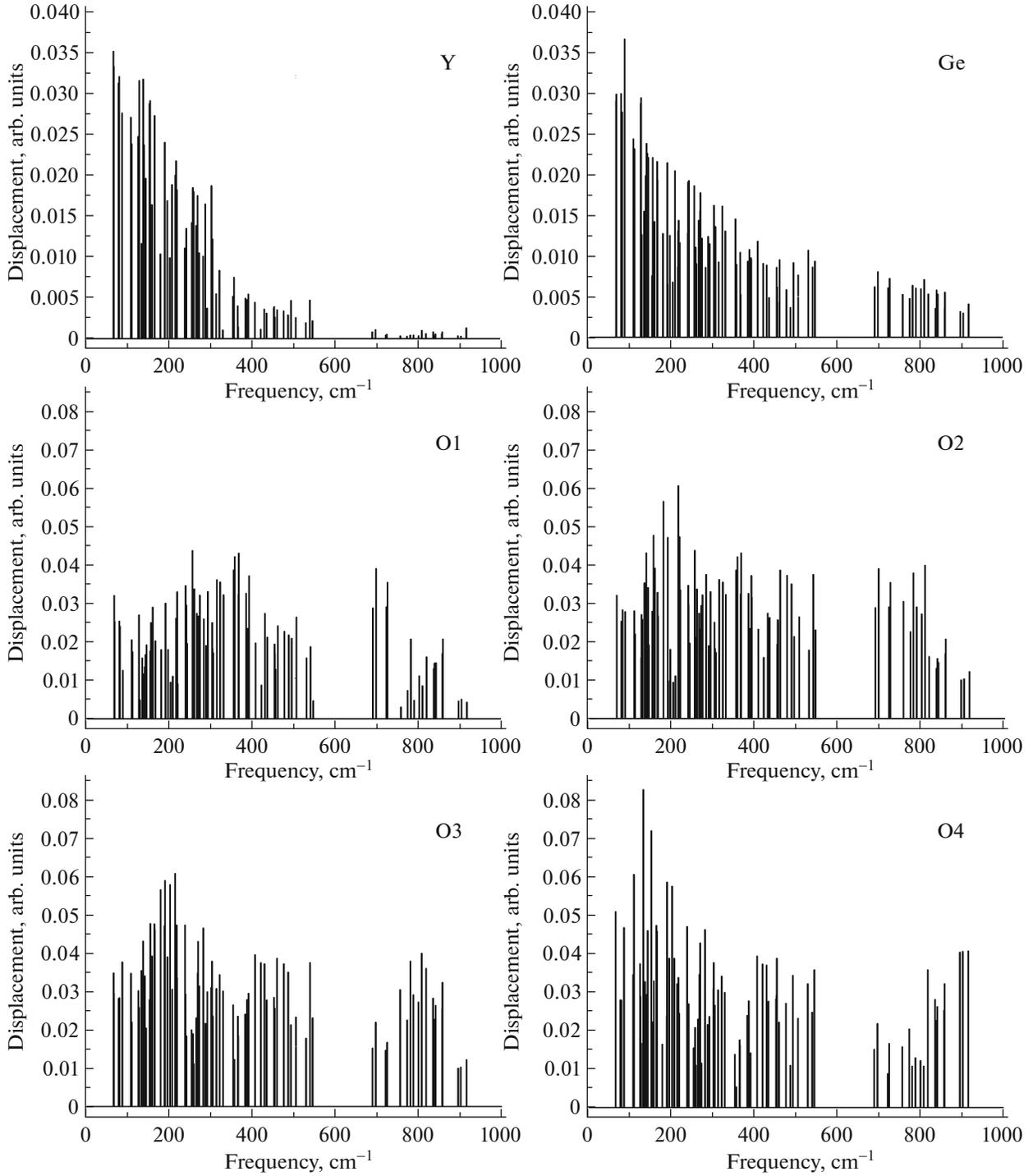

**Fig. 6.** Ionic displacements in the phonon modes for $Y_2Ge_2O_7$.

$$H_V = 0.92\left(\frac{G}{B}\right)^{1.137} G^{0.708}. \quad (1)$$

According to the calculations, all elastic constants in the Tb–Lu series increase with lanthanide compression (except for $C_{12}$, which decreases). The $H_V$ value calculated from formula (1) increases.

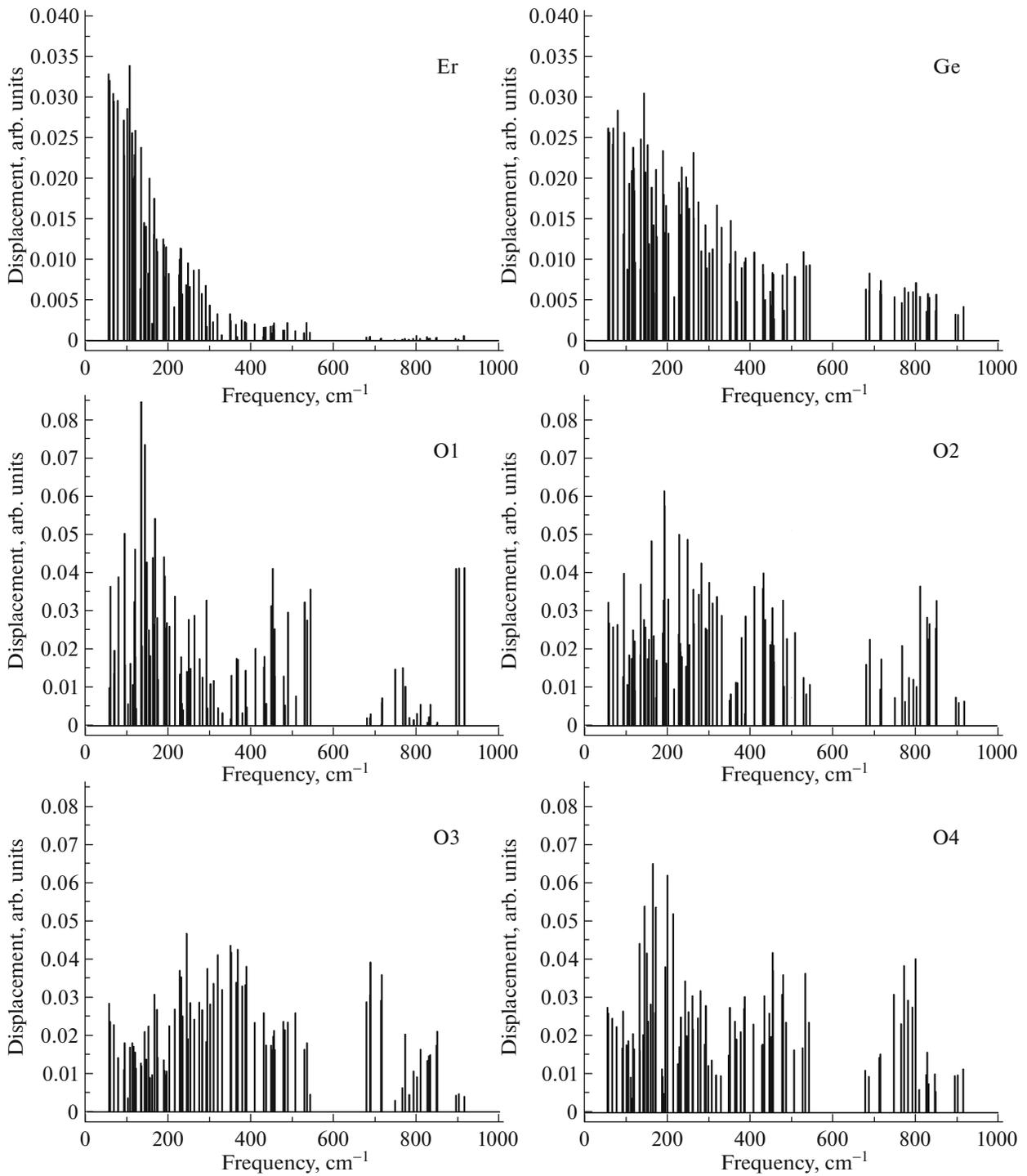

**Fig. 7.** Ionic displacements in the phonon modes for $Er_2Ge_2O_7$.

## 5. CONCLUSIONS

The crystal structures, phonon spectra at the Γ point, and elastic properties of a series of tetragonal germanates $R_2Ge_2O_7$ (R = Tb–Lu, Y) were calculated within the density functional theory and the MO LCAO approach. The frequencies and types of fundamental vibrations and the intensities of the IR and Raman modes were determined. The degrees of participation of ions in each mode were determined by analyzing the displacement vectors obtained as a result of ab initio calculations. The results of measuring the IR and Raman spectra of yttrium germanate $Y_2Ge_2O_7$ were supplemented, and the types of modes were determined; the results obtained can be used to interpret the spectra of isostructural germanates.

**Table 9.** Elastic constants, bulk modulus, shear modulus, and hardness of $R_2Ge_2O_7$ (R = Tb–Lu, Y), GPa

| R | Tb | Dy | Ho | Er | Tm | Yb | Lu | Y |
|---|---|---|---|---|---|---|---|---|
| $C_{11}$ | 251 | 257 | 261 | 266 | 270 | 275 | 278 | 258 |
| $C_{12}$ | 93 | 92 | 91 | 91 | 90 | 90 | 90 | 79 |
| $C_{13}$ | 79 | 80 | 81 | 83 | 84 | 85 | 86 | 76 |
| $C_{33}$ | 268 | 271 | 274 | 277 | 279 | 281 | 284 | 271 |
| $C_{44}$ | 37 | 39 | 41 | 43 | 44 | 46 | 48 | 44 |
| $C_{66}$ | 52 | 54 | 56 | 57 | 58 | 60 | 61 | 59 |
| B | 141 | 143 | 145 | 147 | 148 | 150 | 152 | 139 |
| G | 56 | 58 | 60 | 62 | 63 | 65 | 67 | 63 |
| $H_V$ | 5.50 | 5.80 | 6.08 | 6.35 | 6.60 | 6.86 | 7.05 | 7.02 |

FUNDING

This study was supported by the Ministry of Science and Higher Education of the Russian Federation (project no. FEUZ-2020-0054).

CONFLICT OF INTEREST

The authors declare that they have no conflicts of interest.